\documentclass[12pt]{article}
\usepackage{epsfig}
\usepackage{a4}

\usepackage{a4wide}
\usepackage{wasysym}

\newcommand{\be}{\begin{eqnarray}}
\newcommand{\ee}{\end{eqnarray}}

\newcommand{\ms}{\Delta m^2_{21}}
\newcommand{\ma}{\Delta m^2_{31}}

\newcommand{\sss}{\sin^2 \theta_{12}}
\newcommand{\sch}{\sin^2 \theta_{13}}

\newcommand{\sa}{\sin^2 \theta_{23}}

\def\ltap{\ \raisebox{-.4ex}{\rlap{$\sim$}} \raisebox{.4ex}{$<$}\ }

\begin{document}

\thispagestyle{empty}

\begin{flushright}
{\tt  VEC/PHYSICS/P/2/2005-2006}\\
OUTP-0615P
\end{flushright}

\begin{center}
{\LARGE \bf Modified Zee mass matrix with zero-sum condition\\}
\vspace{2cm}
{\bf Biswajoy Brahmachari$^{\star ^a}$  and Sandhya Choubey$^{\dagger 
b,c}$}
\\
\vskip 1cm
{
$^a${\normalsize \it  Department of Physics, Vidyasagar Evening College,\\
39, Sankar Ghosh Lane, Kolkata 700006, India.\\
$^b${\normalsize \it The Rudolf Peierls Centre for Theoretical Physics,}\\
{\normalsize \it University of Oxford, 1 Keble Road, Oxford, OX1 3NP, UK}\\
$^c${\normalsize \it Harish-Chandra Research Institute,} \\
{\normalsize \it Chhatnag Road, Jhunsi,
Allahabad  211019, India}
}
}
\end{center}
\vskip 1cm
\begin{abstract}

We modify the Zee mass matrix by adding a real one parameter perturbation
which is purely diagonal and trace-less. We show that in this way we
can explain both solar and atmospheric neutrino oscillation
data. There is a correlation between the deviation
from strict maximality of $|U_{\mu 3}|= 1/\sqrt{2}$, with the emergence 
of a small but non-zero $U_{e3}$. We calculate how big a value can 
$U_{e3}$ get when we restrict ourselves within the allowed regions of 
solar and atmospheric neutrino masses and mixing angles. 
We also discuss the impact of a $S_2$ permutation symmetry on 
our mass matrix and show how 
a small $U_{e3} \ne 0$ can emerge when this $S_2$ permutation symmetry
between the second and the third generation is broken. 

\end{abstract}

\vskip 1in

\begin{flushleft}
$^\star$ email: biswajoy.brahmachari@cern.ch\\
$^\dagger$ email: s.choubey1@physics.ox.ac.uk
\end{flushleft}

\newpage

\section{Introduction}

With neutrino oscillations having been successfully established beyond 
all doubts by a series of spectacular experimental results 
\cite{solar,kl,atm,k2k}, the focus has now shifted to 
phenomenologically determining the neutrino mass matrix and 
deciphering the underlying theory which gives us the 
correct neutrino mass matrix. Our current knowledge of the 
range of allowed values for the oscillation parameters at the 
$3\sigma$ limit is \cite{limits}
\be
7.2\times 10^{-5} {\rm eV}^2 < \ms < 9.2\times 10^{-5} {\rm eV}^2 
\label{eq:ms}
\ee
\be
0.25 < \sss < 0.39
\ee
\be
1.4\times 10^{-3} {\rm eV}^2 < \ma < 3.3\times 10^{-3} {\rm eV}^2 
\ee
\be
\sin^22\theta_{23} < 0.9
\ee
\be
\sch < 0.044
\label{eq:sch}
\ee
A series of 
next generation oscillation experiments have been proposed/planned 
to measure very precisely the oscillation parameters. 
Our knowledge on $\ma (\equiv \Delta m^2_{atm})$ and 
$\sa(\equiv \sin^2\theta_{atm})$ 
is expected to improve to about 4.5\% and 20\% accuracy respectively
\cite{huber10}, while that on 
$\ms(\equiv \Delta m^2_\odot)$ and $\sss(\equiv \sin^2\theta_\odot)$ 
could improve to 
7\% and 16\% \cite{th12}. The precision on the mixing angle 
$\sss$ could be improved further with new reactor experiments 
\cite{th12,skgd}.
The hitherto 
unknown mixing angle $\theta_{13}$ will be 
probed in the forthcoming long baseline and 
reactor experiments \cite{white} to values of $\sch \ltap 10^{-3}$ 
\cite{huber10}.
Any hint of CP violation in the lepton sector will be looked for 
in the future long baseline experiments.
The 
deviation of $\theta_{23}$ from maximality and the sign of 
$D_{23} \equiv 0.5 -\sa$ 
can be experimentally checked in 
atmospheric neutrino experiments \cite{th23}. 
The sign of $\ma$ can be probed in either the long baseline experiments
or in atmospheric neutrino experiments \cite{hier}. It might 
even be possible to check the sign of $\ma$ in experiments looking 
for neutrino-less double beta decay ($0\nu\beta\beta$) \cite{0vbbhier}. 
The next generation $0\nu\beta\beta$ experiments are of course expected
to improve the limits on the effective mass parameter to 
$|\langle m \rangle| < 0.03$ eV \cite{0vbb}. 
Limits on the absolute 
neutrino mass coming from direct lab measurement 
is also expected to improve 
from its current limit of $m_\beta < 2.2$ eV to $m_\beta < 0.2$ eV
\cite{katrin}. 
The best current limit 
that we have on the absolute mass scale comes from cosmological 
measurements and these will be further improved in the 
future \cite{cosmonu2006}.
Therefore we expect that the neutrino mass matrix will be determined fairly 
well by the next generation of experiments.

In this 
paper we will phenomenologically analyze a modified version of the 
Zee mass matrix which is compatible with the current allowed values
of the oscillation parameters.
We will study the predictions this model makes for the oscillation
parameters and check how it could be tested in the future.

The Zee ansatz is a theoretically attractive model of neutrino masses and
mixings \cite{zee}. 
This model needs a very minimal extension of the Higgs
content beyond the standard model and one does not necessarily need
to invoke supersymmetry. While the most general mass matrix in the 
Zee model has many more degree of freedoms than can be constrained by 
oscillation data, a more restricted texture for the Zee mass matrix 
emerges if one imposes a condition that only one of the two Higgs 
doublets in the model couples to the charged leptons \cite{zee-wolf}. 
This brings an added advantage that the resultant model does not 
suffer from problems concerning flavor changing neutral currents in 
the charged lepton sector.
This ansatz leads to a neutrino 
mass mass matrix which 
is symmetric and for which the diagonal elements vanish. 
Therefore this mass 
matrix has only three real parameters. 
Consequently, when tested against experimental results, this ansatz
falls in a situation which is
extremely 
constrained \cite{he,zeetest,zeeus}. 
This is because one has to predict two
mass differences and
three mixing angles adjusting only three input parameters. It turns
out that though Zee mass matrix can 
reproduce a maximal $\nu_\mu \leftrightarrow \nu_\tau$ mixing very 
naturally, it fails to simultaneously reproduce the LMA
region of solar neutrino oscillation. 

There has been many attempts to modify the Zee mass 
model \cite{ref3-modif}.
There are mainly three points which are addressed in the literature
which were looked into while Zee model was modified. (i) The
Yukawa couplings which lead to masses of ordinary quarks and leptons
are independent of the Yukawa couplings which generate the neutrino
masses radiatively in Zee mechanism. There were attempts to link these
two families of Yukawa couplings. (ii) There were attempts to embed
the Zee model in grand unified scenarios. (iii) Zee model leads to
severe constraint on $\sin^2 2 \theta_{12} > 0.99$. There
were attempts to modify Zee model which will lead to correct
values of $\sin^2 2 \theta_{12} \sim 0.85$ which is consistent with
the LMA type solution. 

We modify the Zee mass matrix by 
adding a perturbation which is purely diagonal and trace-less.
Furthermore the perturbation has only one real parameter. We
do not introduce any specific field theoretic model for this
extra diagonal elements. This extra piece is introduced in a purely
phenomenological fashion. However, the sum of mass eigenvalues will
remain zero as the mass matrix is trace-less. It has been shown from
very general conditions that a trace-less mass matrix can fit the
observed neutrino oscillation data well \cite{traceless}.

Therefore we consider a real symmetric trace-less neutrino mass matrix with
four parameters. 
We show that this real symmetric and trace-less
four parameter mass matrix 
can correctly predict
the mass squared differences and mixing angles needed to 
explain the world neutrino data. 
Because there
are four input parameters and six testable observables are returned,
one effectively ends up with two predictions.

We begin in section 2 by briefly reviewing the problems faced and 
the current 
status of the original Zee-Wolfenstein ansatz for the mass matrix.
We next analyze 
phenomenologically the mass texture we obtain
by modifying the Zee mass matrix by a trace-less diagonal 
perturbation matrix. We show that this mass matrix can return 
values of neutrino oscillation parameters consistent with the world 
neutrino data. In section 3 we briefly discuss the models which could 
give our neutrino mass texture. We end in section 4 with conclusions.

\section{The ansatz and the phenomenology}

\subsection{The original Zee-Wolfenstein ansatz}

In the Zee model \cite{zee}, 
under the assumption that only one of the Higgs 
doublets couple to the charged fermions \cite{zee-wolf}, the 
neutrino mass matrix assumes the form
\be
{\cal M}_{Zee}=\pmatrix{0 & m_{e \mu} & m_{e \tau} \cr
                 m_{e \mu} & 0 & m_{\mu \tau} \cr
                 m_{e \tau} & m_{\mu \tau} & 0}~,
\label{eq:zee-wolf1}
\ee
which is generally referred to as the Zee mass matrix in the literature.
It can be shown that the three non-zero entries involved can be taken as 
real without any loss of generality and thus the Zee mass matrix
is described by only 3 real parameters.  
Redefining $\sin \theta= {m_{e \mu} \over m_0}$, $\cos \theta={m_{e
\tau} \over m_0}$, and $\epsilon={m_{\mu \tau} \over m_0}$ we get
\be
{\cal M}_{Zee}= m_0 \pmatrix{0 & \sin \theta & \cos \theta \cr
\sin \theta & 0 & \epsilon \cr
\cos \theta & \epsilon & 0}~.
\label{eq:zee-wolf}
\ee
Note that one of the main characteristics of this mass matrix is
its trace-less condition \cite{traceless}. Since the Zee matrix 
is real, this implies that the sum of its 
eigenvalues vanishes,
\be
m_1 + m_2 + m_3 = 0
\ee
This along with the condition that $ee$ element of 
${\cal M}_{Zee}$ is exactly zero, leads one to conclude 
that the only neutrino mass spectrum allowed in this scheme is the 
inverted hierarchy \cite{he}.

If in addition we impose a $L_e-L_\mu-L_\tau$ lepton family symmetry
\cite{bimaxzee,lelmlt}, the entry $\epsilon$ goes to
zero and the resultant matrix 
reproduces bimaximal neutrino mixing as long as 
$\theta=\pi/4${\footnote{
In fact it necessarily produces a maximal solar mixing angle and 
$U_{e3}=0$, while
the atmospheric mixing angle is free to be determined by the value
of $\theta$ which could take any possible value.}.
It also predicts $\ms=0$. To get a non-zero value for $\ms$ one 
has to break the $L_e-L_\mu-L_\tau$ lepton family symmetry slightly,
such that $\epsilon$ is still smaller than the entries 
$\cos\theta$ and $\sin\theta$, with
the difference between the respective magnitudes determined by the 
extent to which $L_e-L_\mu-L_\tau$  is broken. However, even
though this small breaking of the symmetry explains the 
solar mass squared splitting, it fails to drive the solar mixing 
angle far away from maximal mixing as required by the data. 
In fact, up to first order in the small parameter $\epsilon$, 
the solar mass splitting and solar mixing angle is given by
\cite{zeeus}
\be
\ms \simeq 2m_0^2\epsilon \sin2\theta,
\ee
\be
\sss = \frac{1}{2} - \frac{\epsilon}{4}\sin2\theta.
\ee
This implies that the deviation of $\sss$ from maximality 
$D_{12} \equiv 0.5 - \sss$ is given by
\be
D_{12} = \frac{1}{2} - \frac{1}{8}R\sin2\theta,
\ee
where we have used $\ma \simeq m_0^2$. Since from the world neutrino 
data we expect the ratio 
$R\equiv \ms/\ma$ to lie between $[0.022 - 0.066]$, the maximum deviation 
of $\sss$ from its maximal value that we can have in this model 
is 0.00825. This is 
in stark contradiction to the solar neutrino data, which predicts 
a deviation of at 
least $D_{12} >  0.11$ at the $3\sigma$ C.L. In fact, 
the condition $D_{12} < 0.00825$ is ruled out at more than 
$6\sigma$ from the solar 
neutrino data -- meaning 
that the simple Zee-Wolfenstein ansatz is 
disfavored by the data at more than 
$6\sigma$ C.L.

\subsection{The perturbed Zee ansatz}

We add to the mass matrix Eq. (\ref{eq:zee-wolf1}) 
another perturbation matrix which is diagonal and trace-less. 
\begin{equation}
\cal{M}=\pmatrix{0 & m_{e \mu} & m_{e \tau} \cr
                 m_{e \mu} & 0 & m_{\mu \tau} \cr
                 m_{e \tau} & m_{\mu \tau} & 0}
         + \pmatrix{- 2 ~m & 0 & 0 \cr
                     0 & m & 0 \cr
                     0 & 0 & m }~. \label{e1}
\end{equation}
Redefining $\sin \theta= {m_{e \mu} \over m_0}, \cos \theta={m_{e
\tau} \over m_0}, \epsilon={m_{\mu \tau} \over m_0}$ and 
$\delta= {m \over m_0}$ we get,
\begin{equation}
{\cal M}= m_0 \pmatrix{-2 \delta & \sin \theta & \cos \theta \cr
\sin \theta & \delta & \epsilon \cr
\cos \theta & \epsilon & \delta}~.
\label{eq:zee}
\end{equation}

\noindent
An additional $\nu_\mu \leftrightarrow \nu_\tau$ symmetry 
\cite{mutau} 
will force
$\theta= \pi/4$ and simultaneously $\epsilon=\delta$ giving
\be
{\cal M}^\prime= m_0 \pmatrix{-2 \delta & \frac{1}{\sqrt{2}} & 
\frac{1}{\sqrt{2}}\cr
\frac{1}{\sqrt{2}}& \delta & \delta \cr
\frac{1}{\sqrt{2}}& \delta & \delta}~.
\label{eq:zee-mutau}
\ee
This form of the mass matrix has been independently derived from
SU(3) global flavor symmetry \cite{riaz}. However, in this case 
$U_{e3}$ is strictly zero and $\theta_{23}$ is strictly 
maximal\footnote{We will discuss this case again towards the end of
this section.}.
We will work in a scenario where 
$\epsilon$ can be different from 
$\delta$, while both are kept small compared to $\cos\theta$ and 
$\sin\theta$. We will also let $\theta$ to take any possible value.
This would results in $\theta_{23}$ deviating from 
maximal and $U_{e3}$ from zero and would enable us to relate
the deviation of $\theta_{12}$ from its maximal value to explain 
the LMA solution, to the deviation of $\theta_{23}$ and $U_{e3}$ 
from their maximal and null values respectively.

In the limit that $\delta$ and $\epsilon$ are considered small, we 
keep only up to the first order terms in these parameters and 
obtain \footnote{Note that the 
neutrino mass hierarchy predicted in this 
model is inverted and $sgn(\ma)$ is negative.}
\be
\ms \simeq m_0^2 (-2\delta + 2\epsilon \sin2\theta)~,
\label{eq:msol}
\ee
\be
|\ma| \simeq m_0^2~,
\label{eq:matm}
\ee
\be
\sss \simeq \frac{1}{2} - \frac{1}{4}(3\delta + \epsilon \sin2\theta)~,
\ee
\be
\tan^2\theta_{23} \simeq \frac{1}{\tan^2\theta}~,
\label{eq:th23}
\ee
\be
U_{e3} \simeq \sqrt{2}\epsilon \cos2\theta~.
\label{eq:ue3}
\ee
From these expressions we note that the 
deviation of $\theta_{23}$ from its 
maximal value is directly related to the 
deviation of $U_{e3}$ from zero and 
in this model the two are related through
\be
U_{e3} = \sqrt{2}\epsilon \left( \frac{\tan^2\theta_{23} - 1}
{\tan^2\theta_{23} + 1}\right)~.
\ee
Requiring that $0.022<R <0.066$ ($R\equiv \ms/\ma$) 
to satisfy the experimental data, 
we derive the condition using Eqs. (\ref{eq:msol}) and (\ref{eq:matm}) 
\be
0.011 < (-\delta + \epsilon \sin2\theta) < 0.033 ~.
\label{eq:solarmasscond}
\ee
We further note that neither $\theta_{23}$ nor $U_{e3}$ are 
dependent on the value of $\delta$ and depend on $\epsilon$ and 
$\theta$ only. The solar mixing angle on the other hand is given 
from a combination of $\delta$ and $\epsilon$. From the condition 
that $\sss < 0.39$ at $3\sigma$, we derive that 
\be
3\delta + \epsilon \sin2\theta > 0.44~.
\label{eq:solarmixcond}
\ee
One can check that Eqs. (\ref{eq:solarmasscond}) and (\ref{eq:solarmixcond})
are completely compatible. This implies that our 
ansatz for the neutrino mass 
matrix should be completely compatible with world neutrino data.
Combining Eq. (\ref{eq:solarmixcond}) with Eq. (\ref{eq:solarmasscond})
we get the approximate conditions 
\be
 0.107 < \delta < 0.242 ~,
\label{eq:deltarange}
\ee 
\be
\frac{0.119}{\sin2\theta} < \epsilon < \frac{0.275}{\sin2\theta} ~.
\label{eq:epsrange}
\ee
We show in Fig. \ref{fig:param} the values of the parameters
$\epsilon$ and $\delta$ for which our mass matrix predicts 
values for the oscillation parameters consistent within their 
current $3\sigma$ allowed limits 
(given in Eq. (\ref{eq:ms})-(\ref{eq:sch})).
The parameter $\theta$ of the mass matrix is allowed to vary freely 
and we show the results for three different values of $m_0$ given 
in the caption of Fig. \ref{fig:param}. The allowed values for 
$\delta$ and $\epsilon$ roughly are seen to correspond to those given 
by Eqs. (\ref{eq:deltarange}) and (\ref{eq:epsrange}). In Table 1
we give the values of the oscillation parameters predicted by 
our mass matrix for a fixed set of $m_0$, $\theta$, $\epsilon$ and 
$\delta$.

If we denote the deviation of $\theta_{12}$ from its maximal 
value by $D_{12} = 0.5 - \sss$ and deviation of $\theta_{23}$ from 
maximality by $D_{23} = 0.5 - \sa$, then
\be
D_{12} = \frac{1}{4}\left(3\delta + \epsilon \sqrt{1-4D_{23}^2}\right)~,
\ee
while the deviation of $U_{e3}$ from the value of zero is given by
\be
U_{e3} = 2\sqrt{2}\epsilon D_{23}~.
\ee
We show in Fig. \ref{fig:oscparam} the ranges of oscillation 
parameters predicted by our neutrino mass matrix. 
In this figure we fix $m_0^2$ at 0.0015 eV$^2$ (red dots), 
0.002 eV$^2$ (green dots) and 0.003 eV$^2$ (blue dots) and let $\epsilon$,
$\delta$ and $\theta$ take on any possible value such that the 
 world neutrino data is satisfied at the $3\sigma$ C.L. The left 
panels show the correlation between the 
predicted mixing angles while the right
panels give the variation of the ratio $R$ with the 
mixing angles. Note that while the predicted mixing angles and $R$ 
themselves are independent of $m_0^2$, the apparent dependence 
seen in the figure comes from the fact that the mass matrix 
has to simultaneously explain the 
individual mass squared differences $\ms$ and $\ma$, which 
depend on $m_0^2$.

We next turn our attention to the predicted value for the 
effective mass $|\langle m \rangle|$ 
that can be observed in neutrino-less double beta 
decay ($0\nu\beta\beta$) experiments.
Since this is given by just the modulus of the 
$ee$ element of the mass matrix, 
the predicted effective mass in $0\nu\beta\beta$ is
\be
|\langle m \rangle| = 2m_0\delta.
\label{eq:ovbb}
\ee
The prediction for square of the mass observed in 
tritium beta decay experiments is
\be
m_\beta^2 \simeq \frac{m_0^2}{2}\left [ 2 + \epsilon \sin2\theta\right ],
\label{eq:beta}
\ee
while the sum of the masses which is constrained from cosmology is 
predicted in our model to be
\be
\sum_i |m_i| \simeq 2m_0\left[1 + \frac{1}{2}(\delta - 
\epsilon \sin2\theta)\right].
\label{eq:sum}
\ee
The three observables which depend on the absolute value of the 
neutrino masses are related in our model as 
\be
\sum_i |m_i| \simeq \frac{|\langle m \rangle|}{2} - \frac{2m_\beta^2}
{m_0} + 4m_0.
\label{eq:mass}
\ee
Note that while $|\langle m \rangle|$ depends on $\delta$, 
$m_\beta^2$ on $\epsilon$ and $\theta$ and $\sum_i |m_i|$ 
on $\epsilon$, $\delta$ and $\theta$, 
their combination Eq. (\ref{eq:mass})
depends only on $m_0$. 
We show in Fig. \ref{fig:massabs} the predicted values for the 
observables in tritium beta decay ($m_\beta$), $0\nu\beta\beta$
($|\langle m \rangle|$) and cosmology ($\sum_i |m_i|$) in the 
left panels. The right panels show  
the correlation between the effective mass $|\langle m \rangle|$
and the oscillation parameters.

We reiterate that our mass matrix has 
four input parameters namely $m_0$, $\delta$,
$\epsilon$ and $\theta$. The  
overall mass scale as well as the 
mass squared differences are controlled by
$m_0$, whereas it does not influence the 
mixing angles at all. 
Thus $m_0$, which is constrained by the solar and atmospheric 
mass squared differences will have
an upper bound coming from experiments on beta decay,
neutrinoless
double beta decay and cosmology. The mixing angles are 
controlled by the 
three other parameters, namely, 
$\delta$, $\epsilon$ and $\theta$. If we impose the condition 
of maximal $\theta_{23}$ on the mass matrix, then we must  
have $\theta=\pi/4$ (cf. Eq. (\ref{eq:th23})).
If we had a $S_2$ exchange 
symmetry between the second and the third 
generation ($\nu_\mu \leftrightarrow \nu_\tau$), we will get 
$\theta = \pi/4$. However, this symmetry also makes $\epsilon=\delta$
and we get the mass matrix given by Eq. (\ref{eq:zee-mutau}).
The predicted values of the oscillation parameters for this case  
is given by the first row of Table 1. As 
discussed before, in this case the mixing angle
$\theta_{23}$ is exactly maximal, while $U_{e3}$ is exactly zero. 
However, note that the solar mass squared 
difference $\ms$ is also exactly 
zero. That the predicted 
$\theta_{23}$ is maximal, $U_{e3}=0$ and 
$\ms=0$ in this case,
can also be seen from Eqs. (\ref{eq:th23}), (\ref{eq:ue3})
and (\ref{eq:msol}) respectively. 
This mass matrix is therefore phenomenologically untenable. 
In order to be able to explain the neutrino oscillation data, 
this exchange symmetry will have to be broken. 
This case, as 
discussed in detail above, is consistent with all observations. 
The braking of the symmetry forces $\delta \ne \epsilon$ giving 
a small $\Delta m^2_{21}$ and $\theta \ne \pi/4$ causing 
$\theta_{23}$ to deviate from maximality and $U_{e3}$ from zero -- 
all of which are naturally 
small, being protected by the approximate $S_2$ symmetry.

\begin{table}
\begin{tabular}{|c|c|c|c|c|c|c|c|c|c|} \hline
& \multicolumn{3}{c}{Texture parameters}& & $\sin^2 2 \theta_{\odot}$ &
$\sin^2 2 \theta_{atm}$ & $\Delta m^2_{\odot}$
& $\Delta m^2_{atm}$& $U_{e3}$ \\
\cline{2-5}
 & $m_0$ &$\epsilon$ & $\delta$ & $\theta$ &&&(eV$^2$)&(eV$^2$)& \\
\hline
1 & 0.039 & 0.21 &   0.21 & $\pi/4$  & 0.85 & 1.0 
&0 &$1.8 \times 10^{-3}$ & 0 \\
2 & 0.039 & 0.21 &   0.236 & 0.84  & 0.82 & 0.99 
&$9.0 \times 10^{-5}$ & $1.8 \times 10^{-3}$ &$0.019$ \\
3 & 0.039 & 0.21 &   0.236 & 0.73  & 0.82 & 0.99 
&$9.0 \times 10^{-5}$ & $1.8 \times 10^{-3}$ &$0.019$ \\
\hline
\end{tabular}
\caption{ Some specific values of input parameters and outputs. In
figures more details can be found. It can be noted that atmospheric
mixing angle deviates from strict maximality when a non-zero $U_{e3}$
is returned.}
\label{table1}
\end{table}

\section{Discussions on possible models} 

Let us write Eq. (\ref{e1}) as
\begin{equation}
{\cal M}_{ij}=X_{ij} + Y_{ij} \label{e2}
\end{equation}
Here $X_{ij}$ is the symmetric off-diagonal part and 
$Y_{ij}$ is purely diagonal
and trace-less.
In this paper we work with $SU(3)_F$ symmetry as was used in
\cite{riaz} but we consider a further step. We introduce
$SU(3)_F$ symmetry breaking also; which is
obtained by the VEV of an octet Higgs $H_8$
\begin{eqnarray}
&& SU(3)_F {<H_8> \atop\longrightarrow} SU(2)_F \times U(1)_F
\end{eqnarray}
Leptons 
are triplets
and antileptons are antitriplets whereas quarks are singlets. When
$SU(3)_F$ symmetry is broken, electron generation becomes a singlet
whereas second and third generation remains as a doublet of $SU(2)_F$. 
This helps
maximal mixing of atmospheric neutrinos. 
\begin{eqnarray}
\pmatrix{L_e \cr L_\mu \cr L_\tau} \rightarrow L_e + \pmatrix{L_\mu \cr
L_\tau} \\
\pmatrix{e^c \cr \mu^c \cr \tau^c} \rightarrow e^c + \pmatrix{\mu^c \cr
\tau^c} 
\end{eqnarray}
It is easy to see that this symmetry breaking pattern will be obtained
if $H_8$ gets a VEV of the form
\begin{eqnarray}
<H_8> = \pmatrix{-2 & 0 &0 \cr
                 0 & 1 & 0 \cr 
                 0 & 0 & 1} v \label{h8}
\end{eqnarray}
For explaining this let us write down how $SU(3)_F$ 
representations transform under
$SU(2)_F \times U(1)_F$.
\begin{eqnarray}
3 &\longrightarrow &(1,-2)+(2,1) \\
\overline{3} & \longrightarrow & (1,2)+ (2,-1)\\
8 & \longrightarrow &(2,3)+(2,-3) + (1,0) +(3,0)
\end{eqnarray}
Neither $3$ nor $\overline{3}$ can get VEVs because then we will
also break the residual symmetry $SU(2)_F \times U(1)_F$.
The $(1,0)$ component of $8$ must get the VEV; because that
will leave the residual $SU(2)_F \times U(1)_F$ intact. In matrix form the
(1,0) component of the octet $H_8$ is written in Eq. \ref{h8}. 
The $2\times 2$ submatrix in the lower right hand corner is a unit
matrix so this $2 \times 2$ submatrix is a $SU(2)$ singlet. The trace 
is vanishing so $U(1)_F$ quantum number is zero. Thus this specific 
symmetry breaking pattern fixes the form of VEV given in Eq. \ref{h8}.

From the form of the VEV we can see that a residual $SU(2)_F$ symmetry
remains among the second and the third generation which assures
maximality of mixing.
The list of  leptons Higgs scalars and their $SU(3)_F \times SU(3)_c \times
SU(2)_L \times U(1)_Y$ representations are
\begin{eqnarray}
&& FERMIONS \nonumber\\
&& \nonumber\\
&& L = (3,1,2,-1/2) \\
&& l^c =(\overline{3},1,1,1)\\
&& \nonumber\\
&& ZEE~~MODEL~~SCALARS \nonumber\\
&& \nonumber\\
&& \phi_0 = (1,1,2,-1/2) \\
&& \phi_1 = (3,1,2,-1/2) \\
&& \chi = (3,1,1,1) \\
&& \eta = (1,1,2,-1/2)\\
&& \nonumber\\
&& EXTRA~~SCALARS \nonumber\\
&& \nonumber\\
&& \Delta_L = (3,1,3,1) \\
&& H_8 = (8,1,1,0) \\
\end{eqnarray}

First let us explain $Y_{ij}$ term of  Eq. (\ref{e2}).
The Higgs field $\Delta_L$ is a $SU(2)_L \times U(1)_Y$ triplet as
well as a flavor triplet which 
couples  to $\nu_i$ and $\nu_j$ and gives a direct left handed
Majorana mass term of the form $f_{ij} \nu_i \nu_j v_L$, where $v_L$ is
very small. This mass term acts as the diagonal 
and trace-less perturbation if the
Yukawa coupling $f_{ij}$ is diagonal and trace-less.

Let us explain  why the coupling $f_{ij}$ is of this specific form which
is diagonal and trace-less. The reason is that it has its
origin in a higher dimensional operator of the $SU(3)$ flavor group,
\begin{eqnarray}
{f\over M} (3 \times 3 \times 3 \times 8) &=&
{f \over M} ~L~ L~ \Delta_L~ <H_8>
\end{eqnarray}
At this stage the coupling $f$ does not have any generation index. 
Putting the value of $<H_8>$ from Eqn (\ref{h8}) we get the generation
indices as, 
\begin{eqnarray}
f_{ij} L_i L_j \Delta_L
\end{eqnarray}
Now $f_{ij}$ retains the symmetry between the second and third
generation
as well as it is diagonal and trace-less.

In the $X_{ij}$ term of Eq. (\ref{e2}) is symmetric with
diagonal entries vanishing.
It can be generated by
the following diagram in Fig \ref{f1} in a  Zee type model in the presence of
our $SU(3)_F$ flavor symmetry. The VEV of 
$\phi_1(\overline{3})$ is of the order of electroweak scale. This
diagram has been studied very well in the literature. The off diagonal 
mass matrix is of the form
\begin{equation}
m_{i j} \propto g_{i j}~(m^2_j - m^2_i)
\end{equation}
Here $g_{i j}$ is an off-diagonal Yukawa coupling \cite{zee,zee-wolf}. 
Actually $g_{i j}$ controls the coupling strength between two 
lepton doublets and the
charged Higgs $\chi^+$ in Fig \ref{f1}. Therefore $SU(2)_L$
antisymmetry forces it to be off-diagonal. So, when we multiply
$g_{ij}$ by
$(m^2_j-m^2_i)$ the product becomes symmetric with vanishing diagonal
elements. 
%
%
%
%
%
\begin{figure}[h]
\begin{center}
\includegraphics[clip=true, width=14.0cm, height=8.0cm]{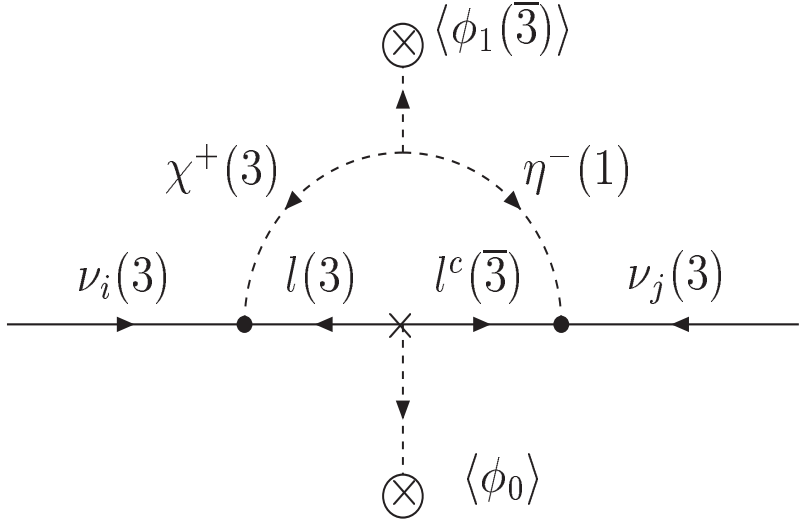}
\end{center}
\caption{\label{f1}
Extra $SU(2)_L$ doublet $\phi_1$ 
transform as $\overline{3}$ of $SU(3)$ flavor. It does not couple to
fermions at tree level.
Quarks do not feel SU(3) flavor in this model. Leptons and
antileptons are triplets and antitriplets of SU(3) flavor. $\chi$ is a
$SU(2)_L$ singlet whereas $\eta$ is a doublet as in minimal Zee model.
In bracket we have displayed the $SU(3)$ flavor quantum numbers.}
\end{figure}

Many different models, in principle,  
can be constructed which are aesthetically more
appealing than this, which will lead to Eq.
(\ref{e1}). We do not know which one is right 
and which is wrong. Typically the first part may be generated by a modified
Zee type mechanism and the second part may be generated by a Higgs
mechanism or a see-saw mechanism or vice-versa. In this paper we do
not focus on the details of the model building apart from citing an
example. More models and
details based on various flavor groups,  will be presented in a future 
paper
\cite{futurepaper}. Here we have added the diagonal and trace-less
perturbation on a
purely phenomenological basis.

\section{Conclusions}

In this paper we have tried to generate simultaneously a non-zero
$U_{e3}$ and LMA solar neutrino mixing angle in a Zee type model by
putting in a diagonal and trace-less perturbation. 
While doing so we have succeeded to keep $\Delta m^2_{21}$ and 
$\Delta m^2_{31}$
as well as the atmospheric mixing angle 
within experimentally allowed ranges.
The resulting mass
matrix texture is symmetric and trace-less. It has four input parameters and
six outputs therefore giving two predictions. When atmospheric neutrino
mixing angle is strictly maximal, $U_{e3}$ vanishes. However when
atmospheric neutrino mixing angle deviates from strict maximality, a
non-vanishing $U_{e3}$ emerges. We studied the correlations between 
the different oscillation parameters as well as observables 
depending on the absolute neutrino mass scale, such as 
the effective mass in neutrino-less double beta decay, 
mass parameter
relevant in beta decay and the sum of the neutrino masses 
relevant for cosmology. The neutrino mass hierarchy is 
predicted to be inverted and the oscillation parameters given 
by this mass matrix are 
well within the reach of future experiments.

\section{Acknowledgements} Work of B.B was financially supported by UGC, New
Delhi, under the grant number F.PSU-075/05-06.

\begin{figure}[t]
\begin{center}
\includegraphics[width=16.0cm, height=16.0cm]{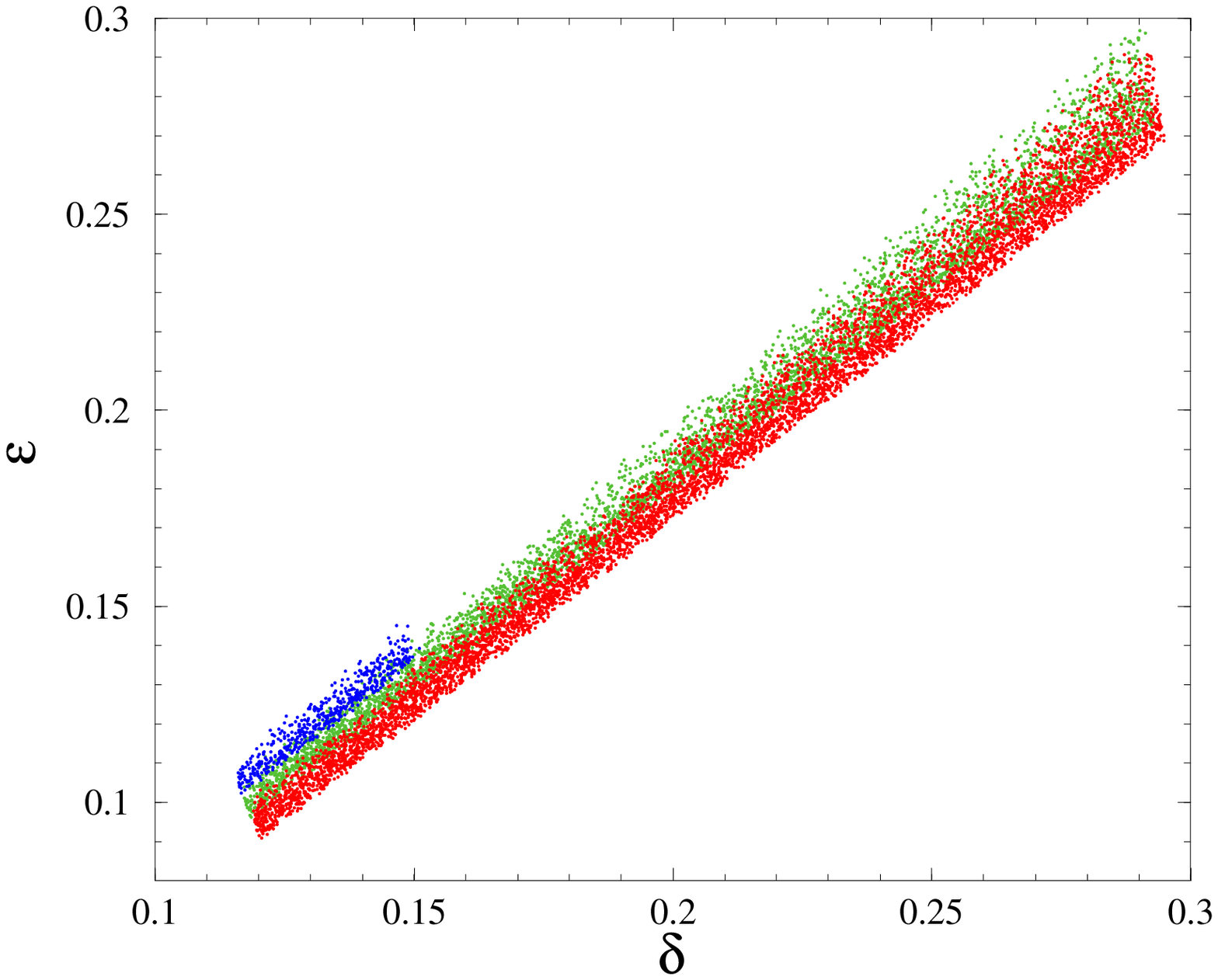}
\caption{\label{fig:param}
The values of $\epsilon$ and $\delta$ which satisfy the current $3\sigma$ 
limits from the world neutrino data. 
The red dots correspond to $m_0^2=0.0015$ eV$^2$, the green dots to
 $m_0^2=0.002$ eV$^2$ and the blue dots to  $m_0^2=0.003$ eV$^2$. 
The parameter $\theta$ of the mass matrix 
is allowed to take any possible value.
}
\end{center}
\end{figure}

\begin{figure}[t]
\begin{center}
\includegraphics[width=16.0cm, height=18.0cm]{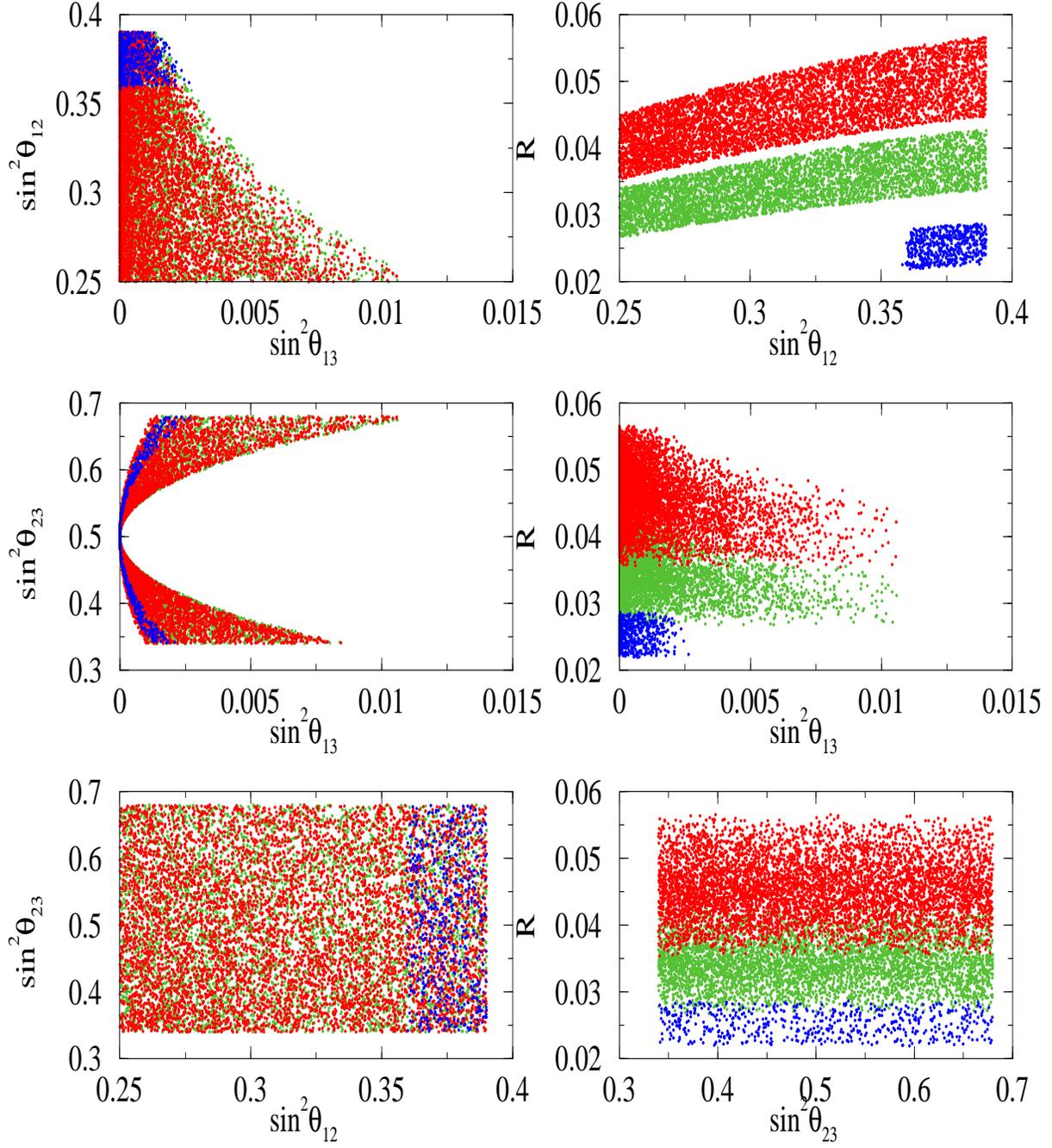}
\caption{\label{fig:oscparam}
Scatter plot showing the correlation between the values of 
the oscillation parameters allowed by our perturbed Zee mass matrix.
The red dots correspond to $m_0^2=0.0015$ eV$^2$, the green dots to
 $m_0^2=0.002$ eV$^2$ and the blue dots to  $m_0^2=0.003$ eV$^2$.
}
\end{center}
\end{figure}

\begin{figure}[t]
\begin{center}
\includegraphics[width=16.0cm, height=18.0cm]{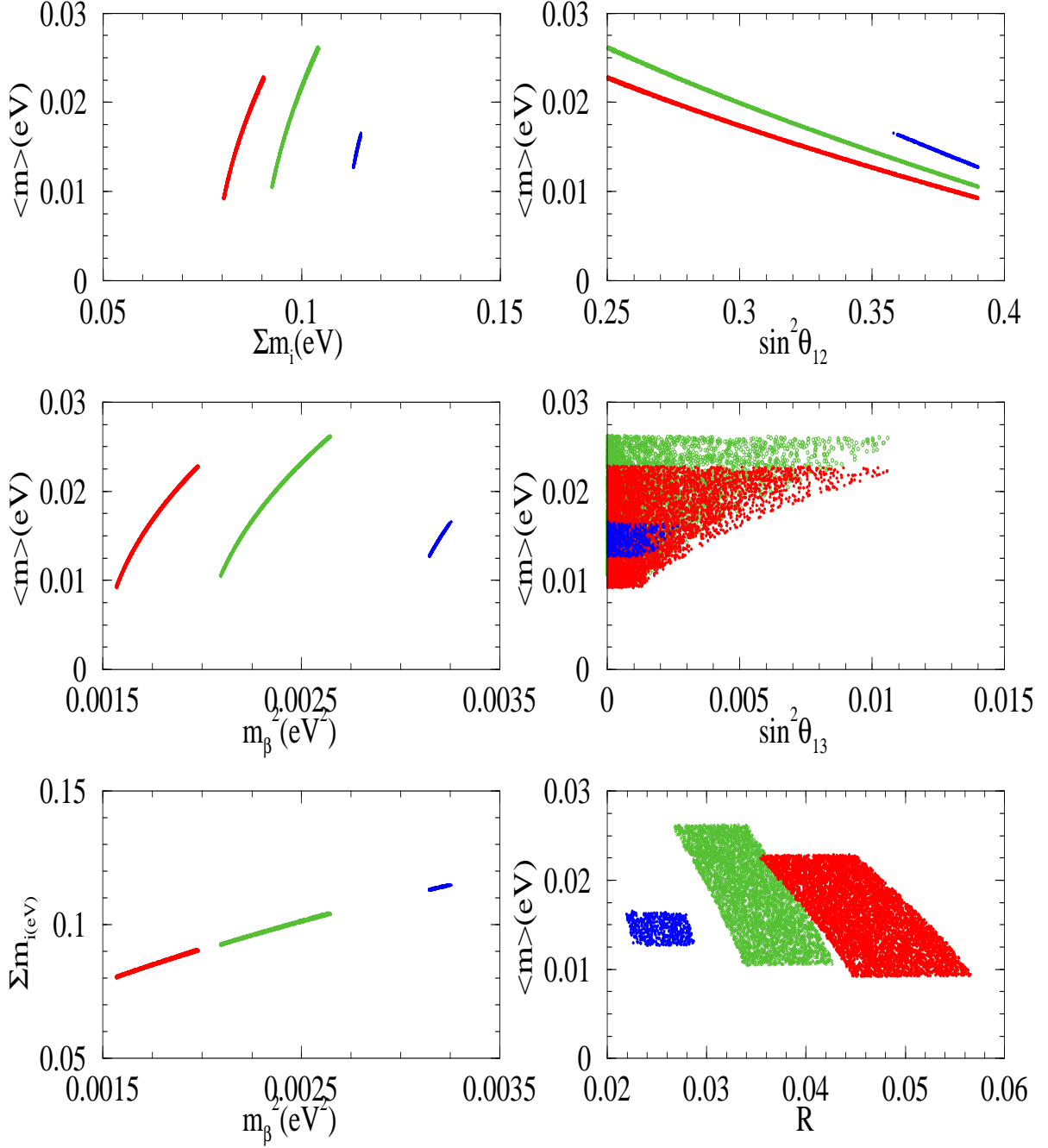}
\caption{\label{fig:massabs}
Left columns give the 
scatter plots showing the correlation between the values of 
$<m>$, $\sum m_i$ and $m_\beta$ allowed by our perturbed Zee mass matrix.
Right columns show the correlations between $<m>$ and the 
mixing angles and ratio $R$.
The red dots correspond to $m_0^2=0.0015$ eV$^2$, the green dots to
 $m_0^2=0.002$ eV$^2$ and the blue dots to  $m_0^2=0.003$ eV$^2$.
}
\end{center}
\end{figure}


\end{document}